\documentclass[aps,prl,twocolumn,superscriptaddress,longbibliography]{revtex4-2}

\usepackage{graphicx}      
\usepackage{amsmath,amssymb} 
\usepackage{siunitx}       
\DeclareSIUnit\angstrom{\text{Å}}
\usepackage{hyperref}      
\hypersetup{
    colorlinks=true,
    linkcolor=blue,
    citecolor=blue,
    urlcolor=blue
}

\newcommand{\Dw}{\Delta\omega}
\newcommand{\dwc}{\delta\omega_{\mathrm{c}}}

\begin{document}


\title{Direct Raman observation of the quantum metric in a quantum magnet}

\author{Chao-Fan Wang}
\thanks{These two authors contribute equally.}
\author{Han Ge}
\thanks{These two authors contribute equally.}
\author{Jun-Yang Chen}
\author{Liusuo Wu}
\email{wuls@sustech.edu.cn}
\affiliation{State Key Laboratory of Quantum Functional Materials, Department of Physics, and Guangdong Basic Research Center of Excellence for Quantum Science, Southern University of Science and Technology (SUSTech), Shenzhen 518055, China}

\author{Xiaobin Chen}
\affiliation{School of Science and State Key Laboratory on Tunable Laser  Technology and Ministry of Industry and Information Technology Key Lab of   Micro-Nano Optoelectronic Information System, Harbin Institute of Technology,  Shenzhen 518055, China }
\author{Jia-Wei Mei}
\email{meijw@sustech.edu.cn}
\author{Mingyuan Huang}
\email{huangmy@sustech.edu.cn}
\affiliation{State Key Laboratory of Quantum Functional Materials, Department of Physics, and Guangdong Basic Research Center of Excellence for Quantum Science, Southern University of Science and Technology (SUSTech), Shenzhen 518055, China}

\date{\today}
\begin{abstract}
The quantum geometric tensor (QGT) unifies the Berry curvature (its imaginary part) and the quantum metric (its real part), yet Raman studies of chiral phonons have so far accessed only the former.  
We perform circularly polarized Raman spectroscopy on the quantum magnet K$_2$Co(SeO$_3$)$_2$, where the field-odd chiral splitting $\Delta\omega$ and the field-even center shift $\delta\omega_c$ collapse onto a single curve across temperature and magnetic field, revealing a common microscopic origin for both observables.  
Since $\Delta\omega$ reflects the Berry curvature, the concomitant even component $\delta\omega_c$, arising from the same microscopic origin, captures the field-induced change of the quantum metric—the diagonal Born–Oppenheimer correction.  
Across two resolvable $E_g$ modes, the unified data are well captured by a simple empirical relation, $\delta\omega_c = \gamma(\Delta\omega)^2$.  
These results establish Raman spectroscopy as a direct probe of the quantum metric and an operational decomposition of quantum geometry within a single measurement.
\end{abstract}
\maketitle

Quantum geometry—comprising the Berry curvature and the quantum metric—governs transport, optics, and superconductivity in quantum materials~\cite{Provost1980, Thouless1982, Xiao2010, Liu2024, Gao2025}. 
Raman studies of chiral phonons have already provided a line-shape handle on the Berry curvature—the imaginary part of the quantum-geometric tensor (QGT)~\cite{Zhang2014, Zhang2015, Zhu2018, Juraschek2025}—but its real part, the quantum metric, has so far lacked a direct spectroscopic observable. 
Consequently, metric effects in solids have been inferred only indirectly from secondary responses such as valley Zeeman shifts, superfluid weight, or nonlinear Hall transport~\cite{Mak2012_NatNano, Srivastava2015, Peotta2015, Sodemann2015, Ma2019_Nature, Sala2025}, or explored in photonic analogs~\cite{Carusotto2013_RMP, Ozawa2019_RMP}.

\begin{figure}[b]
\centering
\includegraphics[width=\columnwidth]{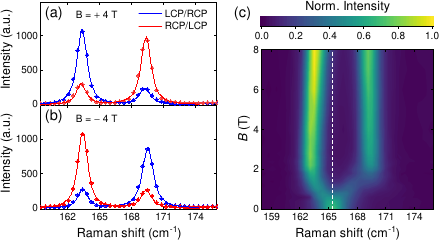} 
\caption{\textbf{Chiral phonon mode P5 at $T=\SI{1.8}{K}$.}
\textbf{a,b,} Helicity-resolved Raman spectra at $B=\pm\SI{4}{T}$ (LCP/RCP in blue, RCP/LCP in red). 
Each cross-helicity channel predominantly excites one branch of the chiral-phonon doublet. 
Finite depolarization gives residual intensity in the opposite channel~\cite{Porto1966}. 
\textbf{c,} Field-dependent normalized intensity map showing the continuous evolution of the two helicity branches, characterized by an odd-in-$B$ $\Delta\omega$ and a concurrent even-in-$B$ $\delta\omega_c$. 
}
\label{fig:1}
\end{figure}

\begin{figure*}[t]
\centering
\includegraphics[width=2\columnwidth]{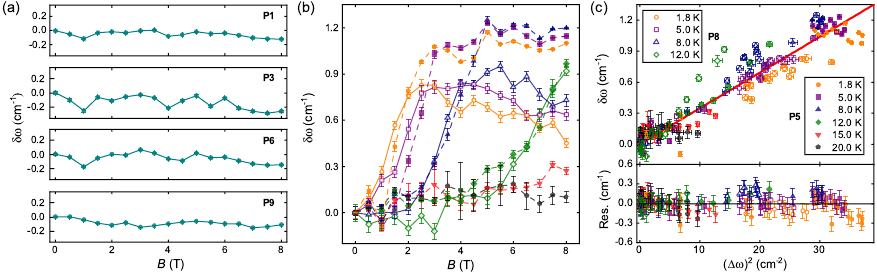}
\caption{
    \textbf{Material-wide quadratic relation linking metric and curvature.}
    \textbf{a,} Nonsplitting reference modes (P1, P3, P6 of $E_g$ and P9 of $A_g$) show negligible center shifts across field, isolating $\dwc$ to the chiral interaction channel.
    \textbf{b,} Mode-dependent center shifts $\dwc(B)$ for the chiral modes P5 (on-site) and P8 (non-local).
    \textbf{c,} Upper: Even-in-$B$ $\dwc$ plotted against $(\Dw)^2$ collapses onto the quadratic law $\dwc = \gamma(\Dw)^2$ with a common slope $\gamma$, unifying the metric and curvature readouts (symbols: P5 solid, P8 empty). Lower: Residuals of the quadratic collapse. Together, these panels demonstrate that both metric and curvature responses originate from a single microscopic interaction channel.
}
\label{fig:universal_law}
\end{figure*}

Here we establish an all-optical, circularly polarized Raman approach that, \emph{within a single measurement}, reveals both components of the quantum-geometric tensor (QGT) as two manifestations of a single origin on the same quantum-magnetic platform.  
A native chiral-phonon doublet serves as an \textit{in-situ} probe whose odd- and even-parity responses arise from the same underlying interaction channel (Fig.~\ref{fig:1}).  
The field-odd chiral splitting $\Delta\omega$ reflects the Berry curvature, whereas the field-even center-frequency shift $\delta\omega_c$ captures the field-induced change of the quantum metric—the diagonal Born–Oppenheimer correction (DBOC)~\cite{Born1956, Mead1979}.  
Their contrasting parity thus provides an operational decomposition of quantum geometry within a single Raman line shape.  
Across temperature and magnetic field, the two observables collapse onto a single master curve (Fig.~\ref{fig:universal_law}), confirming their shared geometric origin.

Experimentally we implement this odd/even Raman decomposition in the triangular-lattice quantum magnet K$_2$Co(SeO$_3$)$_2$ (space group $R\bar{3}m$). 
Magneto-Raman spectra are collected in backscattering, Faraday geometry ($B\parallel c$, $T=1.8$–$300$ K, up to $B=\SI{8}{T}$). 
A representative chiral doublet (mode P5) is shown in Fig.~\ref{fig:1}.  
A full catalog of Raman-active phonons (P1–P9, $\Gamma_{\rm Raman}=4A_g+5E_g$) and crystal-field excitations (CFE1, CFE2) appears in SI Fig.~S2/Table S1\cite{SM_link}, with broader materials context in Refs.~\cite{Zhong2020, Chen2024, Zhu2024, Mou2024}.

Figure~\ref{fig:1} presents the raw helicity-resolved Raman spectra, from which both chiral spitting $\Delta\omega$ and center-frequency shift $\delta\omega_c$ can be identified directly, without any post-processing.  
At $B=\pm\SI{4}{T}$ [Figs.~\ref{fig:1}(a,b)], the two cross-helicity channels (LCP/RCP in blue, RCP/LCP in red) each excite one circular branch of the chiral-phonon doublet, yielding two peak positions $\omega_{+}$ and $\omega_{-}$.  
Their horizontal separation, $\Delta\omega=\omega_{+}-\omega_{-}$, defines the \emph{chiral splitting}, which reverses sign under field reversal (odd in $B$).  
The midpoint $\omega_c=(\omega_{+}+\omega_{-})/2$ marks the \emph{doublet center}; its displacement from the $B{=}0$ reference (white dashed line) gives the \emph{center shift}, $\delta\omega_c=\omega_c(B)-\omega_c(0)$, which is even in $B$.  
Both quantities—$\Delta\omega$ and $\delta\omega_c$—are discernible by eye as the antisymmetric separation and symmetric displacement of the doublet peaks, without any fitting.

The field-dependent intensity map in Fig.~\ref{fig:1}(c) shows the continuous evolution of both branches with magnetic field.  
The antisymmetric splitting and symmetric center shift stand out visually, revealing the two complementary geometric responses—Berry-curvature and quantum-metric components—within a single chiral doublet.  
This observation makes the metric contribution, or diagonal Born–Oppenheimer correction (DBOC), directly accessible as a spectroscopic signal in a crystalline solid.

To understand why these two parities arise from the same underlying mechanism, we turn to the microscopic framework of the phonon self-energy.  
Earlier theory showed that Berry curvature endows phonons with angular momentum and chirality~\cite{Zhang2014, Zhang2015}, and experiments verified the corresponding selection rules~\cite{Zhu2018, Juraschek2025}.  
Recent work extended this concept to include geometric Born effective charges and pseudo-chiral responses in correlated magnets~\cite{Chaudhary2025, Sutcliffe2025}, yet the diagonal Born–Oppenheimer correction—the metric part of a phonon's energy~\cite{Born1956, Mead1979, Xiao2010, Domcke2004_ConicalIntersections, Baer2006_BBO}—has remained inferred rather than observed.  
The following analysis links the two raw observables in Fig.~\ref{fig:1}, $\Delta\omega$ and $\delta\omega_c$, to distinct terms of the same electron–phonon self-energy, establishing the DBOC—and thus the quantum metric—as the even-parity counterpart of the Berry curvature in the Raman line shape.

The odd–even pair in Fig.~\ref{fig:1} arises naturally from the electron–phonon self-energy of a doubly degenerate $E_g$ mode with displacement $\mathbf{u}=(u_x,u_y)$.  
In this minimal picture, the phonon coordinates couple linearly to the electronic degrees of freedom via
\begin{equation}
H_{\text{int}}=\sum_i D_i u_i,\qquad
D_i=\frac{\partial H_e}{\partial u_i}\bigg|_{u=0},
\end{equation}
where $D_i$ denotes the bare electron–phonon operator.  
In the static, long-wavelength limit, these virtual couplings renormalize the phonon potential according to
\begin{equation}
\Pi_{ij}
 =\sum_{n\neq0}\frac{\langle0|D_i|n\rangle\langle n|D_j|0\rangle}{E_0-E_n},
\label{eq:self_energy}
\end{equation}
with $|0\rangle$ and $|n\rangle$ the electronic ground and excited states, respectively.  
Hermiticity imposes $\Pi_{yx}=\Pi_{xy}^*$, and $D_{3d}$ symmetry ensures $\Pi_{xx}=\Pi_{yy}\equiv\Pi_d$.  
The two eigenfrequencies,
\[
\omega_\pm=\omega_0+\delta\omega_\pm
    =\omega_0+\frac{1}{m_\text{eff}\omega_0}(\Pi_d\pm|\Pi_{xy}|),
\]
therefore split and shift according to
\begin{align}
\Delta\omega &= \omega_+-\omega_-=\frac{2|\Pi_{xy}|}{m_\text{eff}\omega_0}, \label{eq:delta_omega}\\
\delta\omega_c &= \frac{\delta\omega_++\delta\omega_-}{2}=\frac{\Pi_d}{m_\text{eff}\omega_0}. \label{eq:delta_omega_c}
\end{align}
Symmetry enforces $\mathrm{Re}\,\Pi_{xy}=0$ and makes $\mathrm{Im}\,\Pi_{xy}$ odd in $B$, immediately producing the parity contrast seen experimentally: $\Delta\omega$ is odd, $\delta\omega_c$ is even\cite{SM_link}.

To connect this self-energy formalism with quantum geometry, we recall that the same virtual matrix elements also construct the electronic quantum-geometric tensor,
\begin{equation}
Q_{ij}
 =\sum_{n\neq0}\frac{\langle0|D_i|n\rangle\langle n|D_j|0\rangle}{(E_n-E_0)^2}
 \;=\; g_{ij}-\frac{i}{2}F_{ij},
\label{eq:qgt_def}
\end{equation}
whose imaginary and real parts correspond to the Berry curvature $F_{ij}$ and quantum metric $g_{ij}$.  
Comparing Eqs.~(\ref{eq:self_energy}) and~(\ref{eq:qgt_def}) shows that both quantities derive from the same set of virtual electronic transitions, differing only by one power of the energy denominator.  
In the single–energy–scale limit (one dominant $\Delta E$), this relation becomes explicit:
\begin{align}
|\Pi_{xy}| &\propto |\mathrm{Im}\,Q_{xy}|\equiv |F_{xy}|=\Omega_e,\\
\Pi_d &\propto \mathrm{Re}\,Q_{xx}\equiv g_{xx}=g_e.
\end{align}
Thus, the two raw observables in Fig.~\ref{fig:1} trace the imaginary and real parts of the same geometric tensor—the Berry curvature and the quantum metric—revealed simultaneously in a single Raman line shape.

Substituting these proportionalities into Eqs.~(\ref{eq:delta_omega})–(\ref{eq:delta_omega_c}) and reinstating the explicit matrix element $|D|^2$ yields
\begin{align}
\Delta\omega &\propto \frac{|D|^2}{m_\text{eff}\omega_0}\,\Omega_e', \label{eq:final_delta_omega}\\
\delta\omega_c &\propto \frac{|D|^2}{m_\text{eff}\omega_0}\,\delta g_e'. \label{eq:final_delta_omega_c}
\end{align}
Here $\Omega_e'$ and $g_e'$ each contain one power of the electronic energy denominator.  
At zero field, $g_e'(0)$ represents the intrinsic quantum metric of the electronic ground state—the conventional diagonal Born–Oppenheimer correction (DBOC), which gives a static frequency renormalization but is normally unobservable.  
The measurable quantity is its field-induced change,
\[
\delta g_e' = g_e'(B) - g_e'(0),
\]
which tracks how the underlying electronic metric evolves under magnetic perturbation.  
In this sense, the magnetic field converts the otherwise static DBOC into a dynamical, spectroscopically accessible response.  
Equations~(\ref{eq:final_delta_omega})–(\ref{eq:final_delta_omega_c}) therefore establish the chiral splitting $\Delta\omega$ and the center shift $\delta\omega_c$ as direct and complementary measures of the Berry curvature and the field-induced change of the quantum metric—the two conjugate components of quantum geometry manifested in one Raman experiment.

\begin{figure}[b]
    \centering
    \includegraphics[width=\columnwidth]{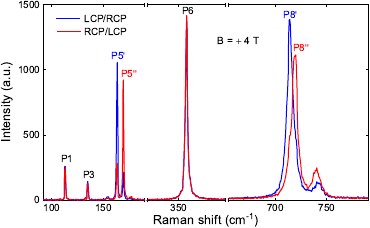}
\caption{\textbf{Mode-selective coupling to quantum geometry.}
Under identical optical and thermal conditions ($T=1.8$~K, $B=4$~T; LCP/RCP and RCP/LCP channels), two $E_g$ modes (P5 and P8) display robust chiral splitting, whereas three $E_g$ modes (P1, P3, P6) remain inert.  
}
\label{fig:2}
\end{figure}
Before detailing the quantitative relation between $\delta\omega_c$ and $\Delta\omega$, we first ask which phonon modes participate in the common geometric channel and which serve as internal controls.  
Fig.~\ref{fig:2} addresses this question by comparing mode-resolved responses under identical optical and thermal conditions ($T=1.8$~K, $B=4$~T, LCP/RCP and RCP/LCP).  
Among the five $E_g$ phonons, only two modes (P5 and P8) exhibit pronounced chiral splitting, while P1, P3, and P6 remain inert.  
This selectivity rules out trivial lattice effects such as strain or crystal-field–induced distortions, which would act indiscriminately on all degenerate modes.  
The inert modes therefore serve as stringent \textit{in situ} controls, confirming that both $\Delta\omega$ and $\delta\omega_c$ stem from specific spin–phonon coupling channels rather than extrinsic shifts.

Equally significant is that the active modes, despite having different absolute frequencies and linewidths, share the same parity behavior: each displays an odd-in-$B$ splitting $\Delta\omega$ accompanied by an even-in-$B$ center shift $\delta\omega_c$.  
This parallel response across disparate phonons reinforces a common microscopic origin tied to the underlying quantum geometry rather than to mode-specific anharmonicities. 
First-principles analysis shows that P5 primarily modulates on-site Co–O bonds, whereas P8 perturbs the extended Co–O–Se–O–Co bridges\cite{SM_link}.  
Distinct atomic-motion patterns thus interrogate different spatial regions of the quantum-geometric tensor (QGT).

\begin{figure}[t]
\centering
\includegraphics[width=\columnwidth]{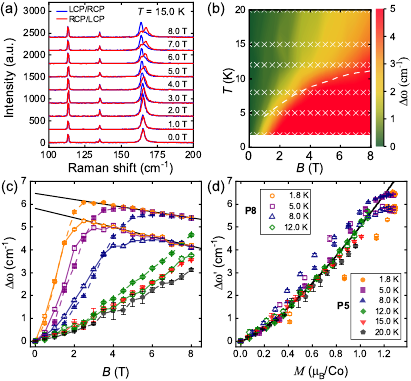}
\caption{
    \textbf{Chiral splitting without LRO and mode-resolved scaling.}
    \textbf{a,b,} P5 (sharp, on-site) retains a robust chiral splitting above $T_{\mathrm{N}}$: raw spectra (\textbf{a}) and a map of $\Dw(B,T)$ (\textbf{b}) both show persistence deep into the paramagnetic regime. The dashed line represents the critical transition temperatures from the paramagnetic state to the $uud$ 1/3-magnetization plateau phase\cite{SM_link}.
    \textbf{c,} $\Dw(B)$ for P5 (solid symbols) and P8 (empty symbols); the $uud$ plateau provides a linear window to extract the Zeeman slopes $\alpha$.
    \textbf{d,} Correlation-driven splitting $\Delta\omega'\equiv\Delta\omega-\alpha B$ versus magnetization $M$ collapses onto $\Delta\omega'=5.03 M^{1.33}$ for both modes prior to the plateau.
}
\label{fig:3}
\end{figure}

Recent studies have reported phonon chirality emerging concurrently with the onset of long-range magnetic order in a variety of magnets, underscoring the diagnostic power of helicity-resolved Raman spectroscopy~\cite{Wu2023, Lujan2024, Yang2025, Che2025, Wu2025}.  Whether long-range order (LRO) is essential remains an open question. As shown in Figs.~\ref{fig:3}(a) and (b), the answer is negative.  
The raw spectra recorded above $T_\mathrm{N}$ [Fig.~\ref{fig:3}(a)] and the corresponding field–temperature map of $\Delta\omega(B,T)$ [Fig.~\ref{fig:3}(b)] both show a robust chiral splitting that persists deep into the paramagnetic regime.  
This persistence demonstrates “chirality without order,” reminiscent of the classic CeF$_3$ behavior~\cite{Schaack1975,Schaack1976,Thalmeier1977}, and shows that the spin–lattice coupling—and hence the underlying geometric response—survives even in the absence of macroscopic order in this triangular magnet.

Figures~\ref{fig:3}(c) and~\ref{fig:3}(d) compare P5 (narrow) and P8 (broad) to quantify how external and internal fields tune the Berry curvature.  
Within the $uud$ 1/3-magnetization plateau at low temperatures, where $M$ is fixed, $\Delta\omega(B)$ varies linearly with $B$; subtracting the Zeeman terms with slopes $\alpha=-0.135(1)$ and $-0.21(1)$~cm$^{-1}$/T for P5 and P8 isolates the correlation-driven component $\Delta\omega'=\Delta\omega-\alpha B$.  
Plotting $\Delta\omega'$ against $M$ yields a common scaling $\Delta\omega' \propto M^{1.33}$ for both modes below the plateau.  
This collapse reveals that $B$ and $M$ act as complementary tuning knobs: the external Zeeman coupling is mode dependent (different $\alpha$), whereas the internal correlation-driven response is nearly identical.

The shared $M^{1.33}$ dependence distinguishes this system from CeF$_3$, where $\Delta\omega\!\propto\!M$ arose from local crystal-field coupling.  
Here, the identical scaling of P5 and P8 implies that the effective curvature arises from correlated $3d$-spin dynamics rather than single-site mechanisms.  
The persistence of $\Delta\omega$ without LRO, together with its field-driven yet mode-independent scaling, indicates that the chiral-phonon response tracks the evolution of the many-body wavefunction—specifically, a field-induced modulation of the quantum metric—rather than simple density-of-states effects.  
This triangular magnet therefore realizes a distinct regime of quantum-geometric spin–lattice coupling, extending the chiral-phonon paradigm beyond the LRO-dependent cases reported previously.

The chiral splitting $\Delta\omega$ observed in circularly polarized Raman spectra is now broadly recognized as a measure of the Berry curvature of the underlying electronic states~\cite{Zhang2014, Zhang2015, Zhu2018, Juraschek2025}.  
Building on this established correspondence, we now examine its even-in-field counterpart—the center-frequency shift $\delta\omega_c$, which arises from the same microscopic origin and encodes the quantum metric.
Figure~\ref{fig:universal_law} demonstrates that these two observables originate from a common microscopic mechanism and manifest a unified geometric behavior, of which the quadratic scaling relation emerges as a direct experimental consequence.

Figure~\ref{fig:universal_law}(a) first confirms selectivity: nonchiral reference modes (P1, P3, P6 of $E_g$ symmetry and the $A_g$ mode P9) exhibit negligible center shifts across all fields, excluding trivial artifacts and isolating $\delta\omega_c$ to the chiral interaction channel.  
Figure~\ref{fig:universal_law}(b) shows the corresponding field-dependence of $\delta\omega_c(B)\equiv\omega_\mathrm{center}(B)-\omega_\mathrm{center}(0)$ for the active chiral modes P5 and P8.  
This apparent complexity simplifies dramatically once $\delta\omega_c$ is plotted against $(\Delta\omega)^2$ in Fig.~\ref{fig:universal_law}(c), which shares the same even-in-$B$ symmetry. All datasets—both chiral modes, all temperatures, and all magnetic fields where two-peak fits remain robust—collapse onto a single curve.

To quantify the connection between the two parity components, we performed a global weighted fit of all 170 data points to the relation
\[
\delta\omega_c = \gamma(\Delta\omega)^2,
\]
using total uncertainties
\[
\sigma_\mathrm{eff}^2 = \sigma_y^2 + (2\gamma\,\Delta\omega\,\sigma_x)^2 + \sigma_\mathrm{sys}^2.
\]
A constant, mode-independent systematic uncertainty of 
\(\sigma_\mathrm{sys}=0.115~\mathrm{cm^{-1}}\)—comparable to the fixed Voigt Gaussian width used in the spectral fitting\cite{SM_link}—was added in quadrature to bring the reduced chi-square to unity (\(\chi^2/\mathrm{ndf}\approx1\)).
This procedure ensures self-consistent weighting across all data sets while leaving the visual error bars (statistical only) unchanged.
The resulting best-fit slope, 
\(\gamma = 0.0346(3)~\mathrm{cm}\),
captures the material-wide proportionality between the metric and curvature responses, validating the single-origin picture.

In summary, a direct, all-optical observation of the quantum metric has been demonstrated, revealing its common microscopic origin with the Berry curvature through helicity-resolved Raman spectroscopy.
By exploiting a native chiral-phonon doublet, we obtain two line-shape observables naturally separated by symmetry: the chiral splitting $\Delta\omega$ (odd in $B$) that tracks the Berry curvature, and the center-frequency shift $\delta\omega_\mathrm{c}$ (even in $B$) that provides direct spectroscopic access to the quantum metric.  
This converts an intrinsic lattice vibration into an \textit{in situ} quantum-geometry sensor that operates without band tuning or external probes.

The same platform further reveals that the underlying spin–lattice coupling—and thus the geometric response—does not rely on long-range order.  
The chiral splitting persists deep into the paramagnetic regime, bridging recent LRO-linked observations with the classic CeF$_3$ case of “chirality without order,” and showing that correlated $3d$ spins sustain finite Berry curvature even when macroscopic order melts.

Most importantly, the metric response follows a universal quadratic law, $\delta\omega_\mathrm{c}=\gamma(\Delta\omega)^2$, characterized by a single material-wide slope $\gamma$.  
This deterministic link between the even- and odd-parity components confirms their common microscopic origin—the real and imaginary parts of one geometric tensor—while elevating the diagonal Born–Oppenheimer correction from a theoretical construct to a measurable spectroscopic parameter.  
In effect, native chiral phonons act as active agents of quantum geometry: they convert geometric “response’’ (Berry curvature) and geometric “fluctuation’’ (metric) into twin observables tunable by field and temperature.

This methodology complements photonic and transport-based probes of quantum geometry~\cite{Ozawa2019_RMP, Carusotto2013_RMP, Ma2019_Nature} and opens a route to engineering correlated solids where curvature and metric can be co-designed.  
Extending this approach to other magnets, multiferroics, and moiré systems may reveal how universal the $\delta\omega_\mathrm{c}$–$(\Delta\omega)^2$ relation is, and whether it can be harnessed for quantum devices that rely on geometric control.  
This unified framework establishes quantum geometry as a directly measurable property of correlated solids.

\begin{acknowledgments} 
We thank Fei Ye for useful discussions. This work was supported by the National Natural Science Foundation of China (Grants No. 12074165, No. 12474143, No. 12134020, and No. 12374146), the National Key Research and Development Program of China (Grant No. 2021YFA1400400), the Shenzhen Fundamental Research Program (Grants No. JCYJ20220818100405013 and No. JCYJ20230807093204010),  the Stable Support Plan Program of Shenzhen Natural Science Fund (Grant No. 20231121101954003), Guangdong Provincial Quantum Science Strategic Initiative (GDZX2401010), the Open Fund of the China Spallation Neutron Source Songshan Lake Science City (Grant No. KFKT2023A06), and the China Postdoctoral Science Foundation (Grant No. 2024M761278).
\end{acknowledgments}

\bibliography{refs}


\end{document}